\documentclass[aps,prb,twocolumn]{revtex4}
\usepackage{amsmath}
\usepackage{epsfig}
\usepackage{graphicx}
\usepackage{multirow}
\usepackage{array}
\usepackage{inputenc}
\usepackage{bm}

\newcommand{\erti}{Er$_2$Ti$_2$O$_7$}
\newcommand{\ersn}{Er$_2$Sn$_2$O$_7$}

\newcommand{\er}{Er$^{3+}$}

\begin{document}
\author{Sylvain Petit$^{1}$, Julien Robert$^{1}$, Sol\`ene Guitteny$^{1}$, Pierre Bonville$^{2}$, 
Claudia Decorse$^{3}$, Jacques Ollivier$^4$, Hannu Mutka$^4$, Michel J.P. Gingras$^{5,6,7}$, Isabelle Mirebeau$^{1}$}
\affiliation{$^1$ CEA, Centre de Saclay, DSM/IRAMIS/ Laboratoire L\'eon Brillouin, F-91191 Gif-sur-Yvette, France}
\affiliation{$^2$ CEA, Centre de Saclay, DSM/IRAMIS/ Service de Physique de l'Etat Condens\'e, F-91191 Gif-Sur-Yvette, France}
\affiliation{$^3$ LPCES, Universit\'e Paris-Sud, 91405, Orsay, France}
\affiliation{$^4$ Institut Laue Langevin, 6 rue Jules Horowitz, BP 156 F-38042 Grenoble, France}
\affiliation{$^5$ Department of Physics and Astronomy, University of Waterloo, Waterloo, Ontario, N2L-3G1, Canada}    
\affiliation{$^6$ Perimeter Institute for Theoretical Physics, 31 Caroline North, Waterloo, Ontario, N2L 2Y5, Canada}
\affiliation{$^7$ Canadian Institute for Advanced Research, 180 Dundas Street West, Suite 1400, Toronto, Ontario, M5G 1Z8, Canada}

\title{Order by Disorder and Energetic Selection of the Ground State in 
the XY Pyrochlore Antiferromagnet \erti. An Inelastic Neutron Scattering Study}
\date{\today}
\begin{abstract}
Examples of materials where an ``order by disorder" mechanism is at play 
to select a particular ground state are scarce. It has recently been proposed, however, 
that the antiferromagnetic XY pyrochlore \erti\, reveals a most convincing case of this 
mechanism. Observation of a spin gap at zone centers has recently been interpreted as a corroboration 
of this physics. In this paper, we argue, however, that the anisotropy generated by 
the interaction-induced admixing between the crystal-field ground and excited levels 
provides for an alternative mechanism. It especially predicts the opening of a spin gap 
of about 15 $\mu$eV, which is of the same order of magnitude as the one observed experimentally.
 We report new high resolution inelastic neutron 
scattering data which can be well understood within this scenario. 
\end{abstract}

\maketitle


Geometrically frustrated magnetism is a forefront research topic within condensed 
matter physics, as testified by the wealth of exotic phenomena discovered over the past years
\cite{Lacroix,gingrasrmp,bramgin00}. For instance, the problem of an XY antiferromagnet on the 
pyrochlore lattice (the celebrated lattice of corner sharing tetrahedra) has been 
 considered with much interest since this model displays an extensive 
classical degeneracy \cite{bramgin94,champ} along with classical and quantum 
order by disorder (ObD) effects \cite{bramgin94,champ,stasiak,zito,savary,gingras1,
gingras2,ludo,zito2,champ2}. The elegant concept of ObD \cite{Villain80,Shender} 
is a cornerstone of ordering in frustrated  condensed matter systems.
 ObD comes into play by selecting a ground state, 
either because fluctuations away from this particular configuration 
allow for a relative gain of entropy compared to other classically degenerate states, or because quantum mechanical zero point fluctuations define a minimum in the total energy. 

Until now, the number of confirmed examples for ObD in real materials have remained scarce \cite{Yildirim}. For ObD to be an efficient selection mechanism, the classical ground state degeneracy must be extremely robust and the minimal theoretical model not openly subject to additional terms that would spoil the accidental emerging symmetry and lift the degeneracy. Recently, the XY pyrochlore antiferromagnet \erti\, has been proposed as a candidate that satisfies these conditions in a rather compelling way \cite{zito,savary,gingras1,gingras2}. Given the unique position of \erti\ among frustrated quantum magnets, it is of foremost importance to scrutinize the soundness of this proposal.

\begin{figure}[t]
\centerline{
\includegraphics[width=8cm]{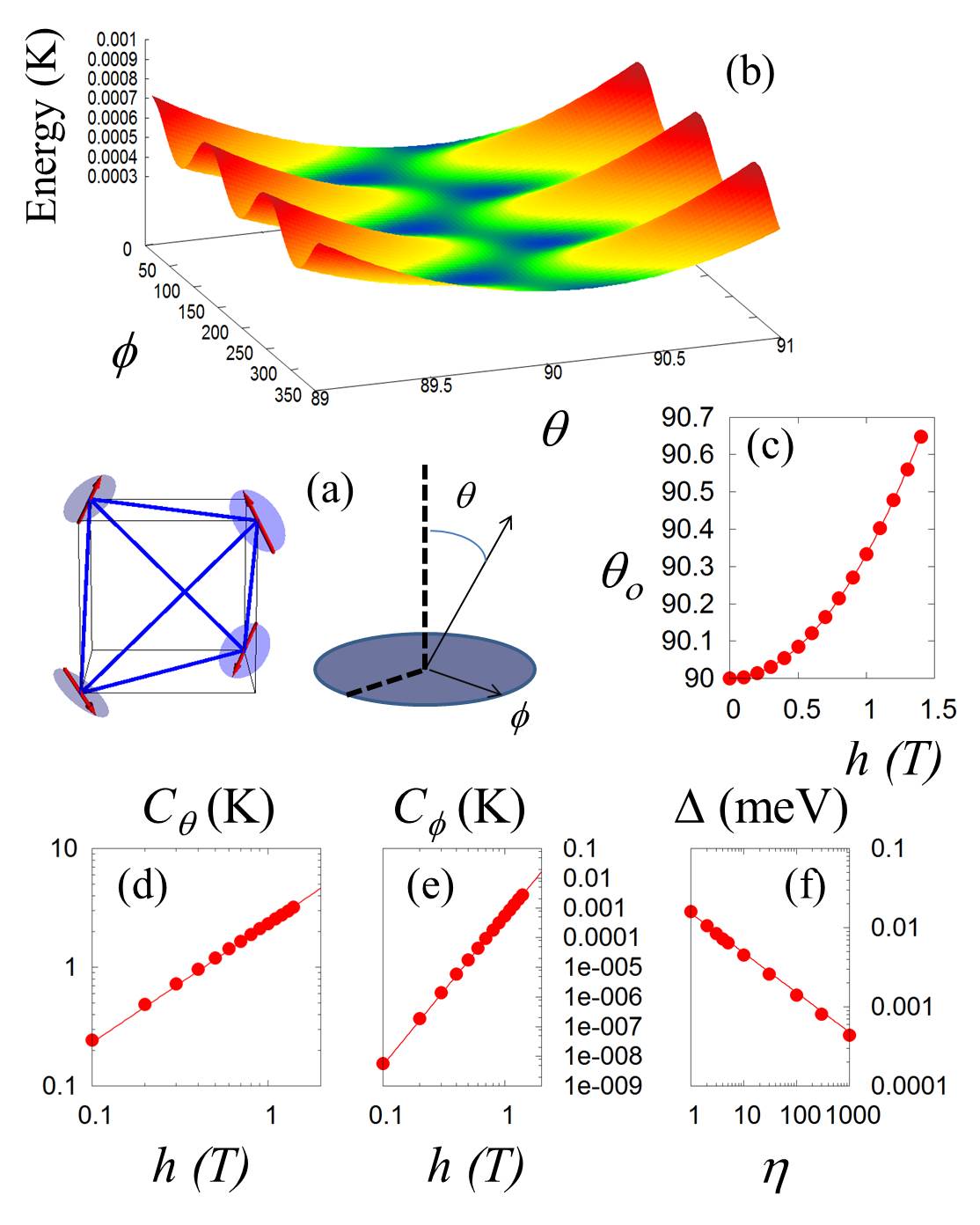}}
\caption{(Color online): (a) Sketch of the $\psi_2$ magnetic configuration. (b) Illustration of the energetic selection by molecular field induced magneto-crystalline anisotropy: the ground state energy of Hamiltonian ${\cal H'}$ computed for $h=1$ T shows minima along the axes $\phi=0, 60, 120, 180$ degrees .., however slightly tilted out of the XY plane ($\theta_o \ne$ 90°). (c) shows the corresponding tilt angle as a function of $h$. (d, e) show respectively the curvature along $\theta$ and $\phi$ of the potential wells as a function of the magnetic field $h$. (f) shows the evolution of the spin gap as a function of the renormalization $\eta$ of the Stevens coefficients $B_{nm}$ (see text).
}
\label{fig1}
\end{figure}

The crystal electric field (CEF) acting on the Kramers \er\, ion is responsible for a 
strong XY-like anisotropy, with easy magnetic planes perpendicular to the local 
$\langle 111 \rangle$ ternary axes \cite{gingrasrmp,champ}. 
Combined with antiferromagnetic interactions, 
an extensive classical degeneracy is expected \cite{bramgin94,champ,stasiak,champ2}.
Despite this degeneracy, \erti\ undergoes 
 a second order phase transition towards an antiferromagnetic 
non-collinear ${\bf k}=0$ N\'eel phase at $T_{\rm N}$=1.2~K \cite{Blote69,Harris98, 
Siddharthan99,champ2}. In this configuration, denoted $\psi_2$ and depicted in Fig. 
\ref{fig1}(a), the magnetic moments are perpendicular to the $\langle 111 \rangle$ 
 axes \cite{champ2,poole} and make a zero net magnetic moment per tetrahedron.

A theory based on a Hamiltonian written in terms of interacting pseudospins 1/2, each describing
the single-ion CEF ground doublet, along with four anisotropic nearest-neighbor exchange parameters 
$({\sf J}_{\pm\pm}, {\sf J}_{\pm}, {\sf J}_{z\pm},{\sf J}_{zz})$, 
has been proposed for \erti\ \cite{savary}.
For the set of parameters determined by inelastic neutron scattering (INS) experiments 
in a large applied magnetic field \cite{ruff,savary}, the theory \cite{savary} predicts a quantum ObD 
selection of $\psi_2$, on the basis of a linear spin wave calculation \cite{stasiak,zito,
savary,gingras1}, as well as thermal ObD at $T_c$, also selecting $\psi_2$ \cite{gingras2}. 
Another consequence of ObD in \erti\ is the opening of a spin gap, previously 
inferred from EPR experiments \cite{sosin} as well as from deviation of the $T^3$ law 
in specific heat measurements \cite{reotier}, and very recently confirmed from INS measurements \cite{ross}.
However, while the spin gap is a necessary consequence, it is not a definitive proof of this 
scenario: whatever the mechanism, a spin gap is expected since the ordered $\psi_2$ 
ground state breaks a global discrete symmetry \cite{stasiak}. 

In this work, we follow a different route and consider an anisotropic bilinear exchange Hamiltonian written for the \er\, moments along with the CEF contribution (henceforth referred to as Model A). As shown in Ref. $[$\onlinecite{clarty}$]$, an energetic selection of the $\psi_2$ state is possible for a specific range of anisotropic exchange parameters, owing to magneto-crystalline effects described by the CEF (see also Ref. $[$\onlinecite{ersn}$]$). We model the spin excitations spectra within this CEF-induced energetic selection scenario. The comparison with new INS data allows one to determine a new set of anisotropic exchange parameters. These are compatible with those determined from in-field INS experiments \cite{savary} and based on the pseudospin 1/2 Hamiltonian (referred to as Model B), from which quantum \cite{zito,savary,gingras1} and thermal (at $T_{\rm N}$) ObD \cite{gingras2} is predicted. Both approaches lead to a spin gap of the correct order of magnitude. However, in model A, the spin gap results strictly from the admixing of the CEF levels via the mean-field \cite{clarty}. Our results revive the debate regarding the ordering mechanism in pyrochlore antiferromagnets and illustrate that the argument of quantum ObD being the chief governing mechanism causing $\psi_2$ ordering in \erti\, is not definitive. More generally, they emphasize the limitations of the projection onto the pseudospin 1/2 subspace (shift from model A to model B) with solely bilinear anisotropic spin-spin coupling in describing even qualitatively the physics of highly frustrated rare-earth pyrochlores.


\begin{figure*}[t]
\centerline{
\includegraphics[width=18cm]{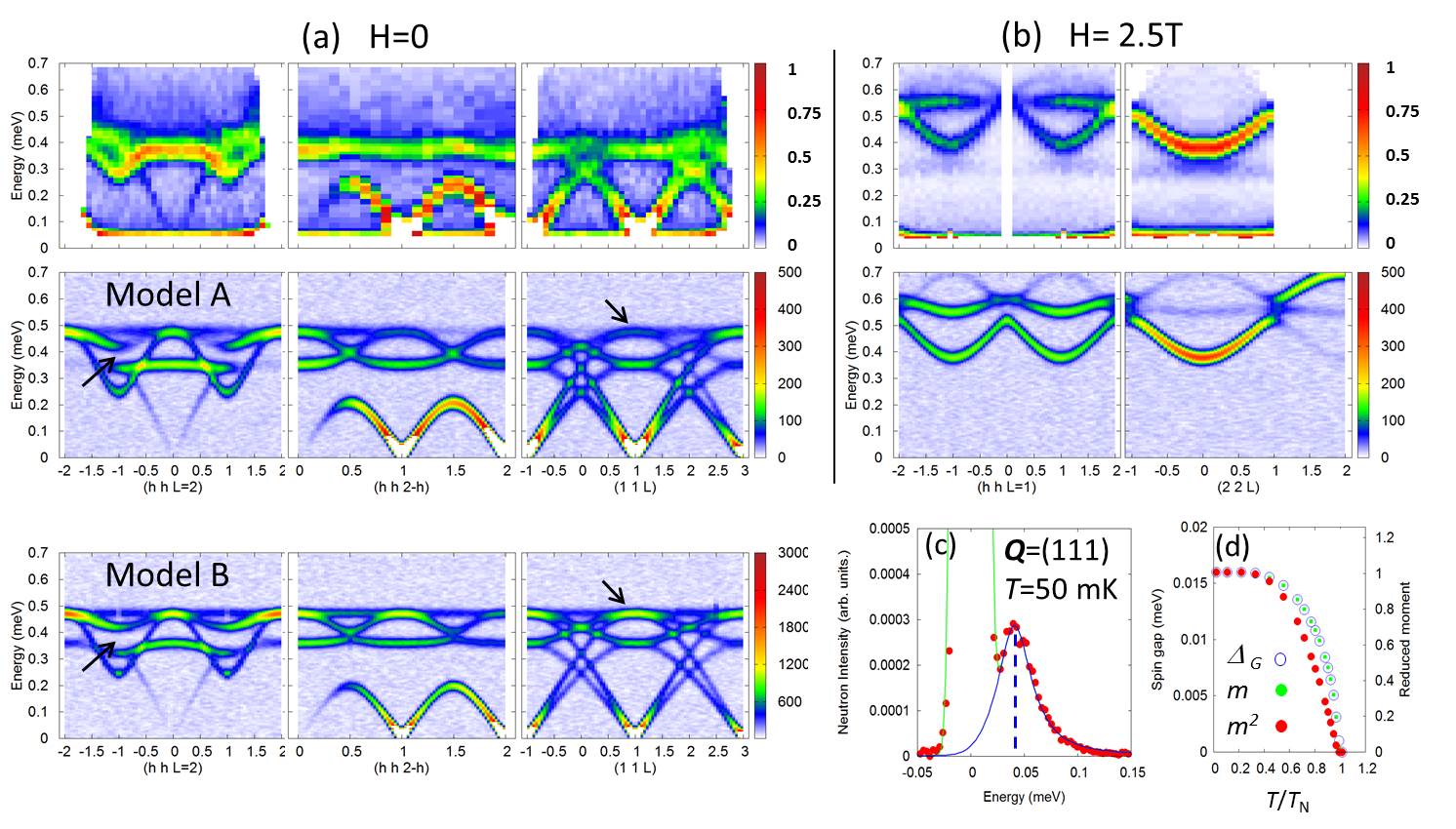}}
\caption{(Color online): IN5 time of flight spectra taken at 50 mK with an incident wavelength of 
6 \AA\, along various directions, in zero field (a) or under a magnetic field of 2.5 T applied along 
the $(1,-1,0)$ direction (b). $S(Q,\omega)$ is shown for the RPA (Model A) and spin half (Model B), 
taking into account equi-populated $\psi_2$ domains, as described in the main text. The global 
agreement is evident but the arrows point out specific Q regions showing the limits of the two models. 
(c) INS raw data recorded with an incident wavelength of 8.5 \AA\ at ${\bm Q}=$(111) and $T=$50 mK showing the spin gap (blue line) at 
43 $\mu$eV. (d) Evolution of the spin gap and of the magnetic moment in Model A as a function of temperature. The saturated moment is 3.7 $\mu_B$ while $T_{\rm N}$=2.1 K (at the mean field level).
}
\label{fig-tof}
\end{figure*}

{\it CEF energetic selection mechanism (Model A)} $-$ 
This approach considers a mean-field anisotropic bilinear exchange Hamiltonian 
written for the \er\, moments $\vec{J}_i$ at sites $i$ of the pyrochlore lattice (see Refs.
$[$\onlinecite{clarty}$]$ and $[$\onlinecite{ersn}$]$). It also contains explicitly the 
CEF contribution, ${\cal H}_{\mbox{CEF}}$:
\begin{equation}
{\cal H} = {\cal H}_{\mbox{CEF}} + \frac{1}{2}\sum_{i,j} \vec{J}_i \cdot {\cal K}_{i,j} \cdot \vec{J}_j 
\label{eqmodela}
\end{equation}
where ${\cal H}_{\mbox{CEF}}=\sum_{n,m} B_{nm} O_{nm}$ is written in terms of $O_{nm}$ Stevens operators \cite{wybourne,jensen}. The $B_{nm}$ have been determined to fit a number of experiments, including the intensities and positions of the crystal field levels, as well as the susceptibility \cite{cao,bnm,supmat}. In the following, those values are 
considered as fixed parameters. ${\cal K}_{i,j}$ denotes an anisotropic coupling tensor, 
defined in the $(\vec{a},\vec{b},\vec{c})$ frame attached to the \er$-$\er\, bonds \cite{malkin,supmat}. 
It is described by 3 symmetric parameters, ${\cal K}_{a,b,c}$, 
and an anti-symmetric exchange constant (Dzyaloshinskii-Moriya like), ${\cal K}_4$ \cite{supmat}.

In this model, the molecular field induces an admixture between the ground 
and excited CEF levels, leading to an effective magnetic anisotropy. This point 
is best evinced by considering the problem of an \er\, ion in a local magnetic 
field $\vec{h}_i$: 
${\cal H}' = {\cal H}_{\mbox{CEF}} + g_J \mu_B \vec{h}_i \cdot \vec{J}_i$. 
Figure \ref{fig1}(b) shows the ground state energy of ${\cal H}' $ computed 
as a function of $\theta$ and $\phi$ (in the local basis) for a field $h=1$ T, which 
is the actual order of magnitude of the molecular field in \erti\ (see below). Minima along the 
6-fold directions of the CEF, {\it slightly tilted} away from the XY plane perpendicular to the local
$[111]$ direction, are clearly observed. 
As shown in Fig. \ref{fig1}(c), the tilt grows as $h^2$ but remains less that 
one degree for realistic values of $h$. The potential well in the vicinity of the 
minima can be approximated by a highly anisotropic harmonic potential whose 
curvature along $\theta$ (denoted by $C_{\theta}$) is far steeper than along 
$\phi$ (denoted by $C_{\phi}$). The average curvature, 
given by $\sqrt{C_{\theta} C_{\phi}}$ \cite{jensen}, is approximately $3\times 10^{-2}$ K 
at $h=1$ T, a value about the same order of magnitude as the one emerging 
from zero point fluctuations (quantum ObD) \cite{savary}. 

Returning to the model A Eq. (\ref{eqmodela}), this effective anisotropy combined 
with appropriate exchange parameters stabilizes the $\psi_2$ state (note that other 
phases, namely a canted ferromagnet as well as the antiferromagnetic Palmer and 
Chalker state can also be stabilized depending on the values of ${\cal K}_{a,b,c,4}$ 
\cite{ersn}). Here, the moment direction at each of the 4 sites of a primitive tetrahedron basis 
is given by $(x,x,y), (-x,-x,y), (-x,x,-y), (x,-x,-y)$ with $y \approx 2x$ (in the cubic frame). 
Note that this is allowed by symmetry for the $\psi_2$ state, but not for the other 
component ($\psi_3$) of the $\Gamma_5$ two-dimensional representation \cite{javanparast}. 

The dynamical structure factor $S({\bm Q},\omega)$ that exposes the spin dynamics is modeled 
by a random phase approximation (RPA) 
 calculation (see Refs $[$\onlinecite{jensen,kao}$]$ and Supplemental Material 
\cite{supmat}). For the relevant set of ${\cal K}_{a,b,c,4}$ parameters (see below), 
numerical calculations show the opening of a spin gap $\Delta_G^{\rm RPA} \approx 15$ $\mu$eV at Brillouin zone centers. 
To emphasize explicitly the influence of the CEF levels in causing this gap,
 calculations have been performed for $B_{nm}$ parameters multiplied by a renormalization coefficient $\eta$. This has the effect of rescaling all CEF energy gaps by $\eta$. In these calculations, the exchange coupling remains fixed, so that the molecular field and the N\'eel temperature are essentially unchanged, but the effective magnetic anisotropy inherited by admixing of the ground CEF doublet with excited CEF states decreases with increasing $\eta$. $S({\bm Q},\omega)$ is also globally unchanged, yet the spin gap gradually decreases and tends to zero for large $\eta$, Numerical calculations show that $C_{\theta} ~\sim h$ while $C_{\phi} ~\sim h^5/\eta^4$. Ultimately, as $\eta \rightarrow \infty$, the $U(1)$ classical ground state degeneracy within the Hilbert space strictly composed of of a direct product of single-ion CEF ground doublets is recovered (see Fig. \ref{fig1}f). In that limit, quantum \cite{zito,savary,gingras1} 
and thermal \cite{gingras2} ObD would become the sole mechanism able to lift the accidental degeneracy.

{\it Comparison with experiments} $-$ To determine the ${\cal K}_{a,b,c,4}$ couplings, the INS data were fitted to the calculated $S({\bm Q},\omega)$ within the RPA. The neutron measurements were performed on a large \erti\, single crystal grown with the floating zone technique. The crystal was inserted in a copper sample holder and attached on the cold finger of a dilution fridge, allowing one to cool the sample down to 50 mK. Data were collected on the IN5 time-of-flight instrument (ILL) which combines high flux with position sensitive detectors allowing for single crystal spectroscopy. Measurements were carried out with an incident neutron wavelength of 6 \AA in zero field and under an applied magnetic field of 1.5 and 2.5 T along $[1,-1,0]$. The spin excitation spectra measured along the high symmetry directions of the cubic unit cell at 50 mK are shown in Fig. \ref{fig-tof}. These results compare well with prior measurements (see Ref.~[\onlinecite{ruff}] and the supplemental material of Ref.~[\onlinecite{savary}]). Because of the magnetic $\psi_2$ domains, the identification of the expected four different spin wave branches is not straightforward. This means that the inelastic peaks in Fig. \ref{fig-tof} contain several modes within the experimental resolution. This is evidenced in the high resolution set-up, using a wavelength of 8.5 \AA. The highly Gaussian (nearly triangular) profile of the resolution line inherent to the counter-rotating disk choppers instrument, which yields an energy resolution of about 20 $\mu$eV, reveals two acoustic-like modes coming from different magnetic domains (see  Supplemental Material). Using the same high resolution set-up, the zero field data confirm the opening of a spin gap at zone centers: as shown in Fig. \ref{fig-tof}(c), the energy resolution permits to discriminate between the inelastic scattering and the strong Bragg intensity at the ${\bm Q}=$ (111) position. Above the elastic line, the neutron intensity first shows a dip and then a peak, a behavior that is typical of a spin gap. Fitting the data through a Lorentzian profile convoluted with the resolution function (see blue line in Fig. \ref{fig-tof}c) yields $\Delta_{G}^{{\rm exp}} \approx 43$ $\mu$eV. This value compares very well with previous estimates \cite{sosin,reotier,ross}. 

To determine the exchange parameters, we calculate $S({\bm Q},\omega)$ assuming 
an equal population of the six $\psi_2$ magnetic domains. On the basis of exhaustive calculations 
as a function of the parameters ${\cal K}_{a,b,c,4}$ in zero and applied magnetic field, 
the INS data were fitted by matching the location of the maximum INS intensity in 
several directions. A good agreement is found for the following values: 
\begin{eqnarray}
{\cal K}_a & \sim & 0.003 \pm 0.005~{\rm K} \quad {\cal K}_b \sim 0.075 \pm 0.005 ~{\rm K} \nonumber \\
{\cal K}_c & \sim & 0.034 \pm 0.005~{\rm K} \quad {\cal K}_4 \sim 0 \pm 0.005 ~{\rm K} .
\label{Kresults}
\end{eqnarray}

Many others sets that capture independently the magnetization or the excitation spectrum 
can be found but the present determination provides values that capture all these experimental data \cite{supmat}. 
With these exchange parameters, the spin gap is evaluated at $\Delta_{G}^{\rm RPA} \approx$ 15 
$\mu$eV, a value smaller than $\Delta_{G}^{{\rm exp}}$, but of the correct order of magnitude. 

\begin{table}[t]
\begin{tabular}{*{4}{c}}
\hline
Coupling & Model A   &Model B& Ref. $[$\onlinecite{savary}$]$ \\ 
\hline
${\sf J}_{\pm \pm}$ & 4.54 ($\pm$ 0.1) & 4.3 ($\pm$ 0.1) & 4.2 ($\pm$ 0.5)  \\
${\sf J}_{\pm}$   & 5.84 ($\pm$ 0.1) & 6.0 ($\pm$ 0.1) & 6.5 ($\pm$ 0.75) \\
${\sf J}_{z\pm}$   & 0.92 ($\pm$ 0.1) & -1.5 ($\pm$ 0.1) & -0.88 ($\pm$ 1.5) \\
${\sf J}_{zz}$    & -0.87 ($\pm$ 0.1) & -2.2 ($\pm$ 0.1) & -2.5 ($\pm$ 1.8) \\
\hline
\end{tabular}
\caption{Anisotropic exchange parameters. Units are in $10^{-2}$ meV. Positive 
values correspond to AF interactions.}
\label{table1}
\end{table}

{\it Discussion} $-$ It is instructive to compare our results (Eq.~(\ref{Kresults})) with those found using the 
pseudospin $1/2$ approach (Model B) \cite{savary}. The corresponding anisotropic exchange 
Hamiltonian is described in detail in Ref. $[$\onlinecite{savary}$]$ and is 
based on the anisotropic couplings $({\sf J}_{\pm\pm}, {\sf J}_{\pm},{\sf J}_{z\pm},
{\sf J}_{zz})$ acting between pseudospin $1/2$ components written in their local basis. 
The calculation of the dynamical structure factor for this model
was performed within the Holstein-Primakov approximation, using 
the {\it Spinwave} software developed at the LLB \cite{spinwave}. Following the same 
fitting procedure as above, a set of parameters is obtained which largely confirm the results of 
Ref. $[$\onlinecite{savary}$]$ (see Table \ref{table1}\footnote{The values reported in the first column are different from those found in \cite{petitprbrc}. The present values correct a numerical error in translating the couplings from meV to K. The calculations as performed and reported in \cite{petitprbrc} were done however with the correct numerical values}). 
The most striking point is that 
Models A and B lead to very similar $S({\bm Q},\omega)$. This is due to the fact that 
both models adopt a predominant effective Hamiltonian with bilinear couplings in 
terms of pseudospin $1/2$ operators when projected in the CEF 
ground doublet \cite{thompson}. 
Specifically, there is a relationship \cite{ersn} between the two sets of anisotropic 
exchange couplings based on the $g_{\perp}$ and $g_{z}$ Land\'e factors deduced 
from the ground state doublet wavefunctions \cite{supmat}. Table \ref{table1}, 
which allows to compare ${\cal K}_{a,b,c,4}$ transformed in the 
$({\sf J}_{\pm\pm}, {\sf J}_{\pm},{\sf J}_{z\pm},{\sf J}_{zz})$ language with the 
values determined from model B and from Ref. $[$\onlinecite{savary}$]$, shows that 
the sets of values are similar whether determined from either model. Further, owing to 
the ObD mechanism, Model B leads to a spin gap $\Delta_G$= 21 $\mu$eV 
\cite {savary}, a value of the same order of magnitude as the one 
($\Delta_G^{\rm RPA}$ = 15 $\mu$eV) obtained in Model A.

While the maps in Fig.~\ref{fig-tof} demonstrate an overall agreement with experiment, 
some discrepancies are observed nonetheless, which equally affect Models A and B. 
The most important difference concerns the acoustic-like mode stemming from (0,0,2), which 
seems to disperse continuously up to 0.45 meV in the neutron data. Within the experimental 
uncertainty, there is no gap opening when this branch crosses the optical one (see the arrows in 
Fig. \ref{fig-tof}(a), left and right columns of the $H$=0 panel). Such a gap opening occurs in the calculations, 
separating the acoustic branch from a higher energy optical branch. Furthermore, both models 
predict two well-separated modes at the zone centers ${\bm Q}$ = (1,1,1), (2,2,0) and (0,0,2) 
at about 0.45 and 0.5 meV, whereas a single one is observed in experiment (middle column of 
the $H$=0 panel in Fig. \ref{fig-tof}(a)). More elaborate models are probably necessary to explain 
these features, taking into account the long-range part of the dipolar interaction or more complex 
coupling terms than bilinear ones.

Reference $[$\onlinecite{ross}$]$ reports the evolution of the gap as a function 
of temperature, and ascertains that it varies as the square of the $\psi_2$ order parameter. The gap 
calculated in the framework of Model A shows instead a linear evolution with the 
order parameter (see Fig.~\ref{fig-tof}(d)). The difference between the linear 
and squared order variations is most pronounced in a narrow temperature range 
spanning $T_N$ to $0.6 \times T_N$. Unfortunately, in this temperature range, 
we believe that the experimental uncertainty in Ref. $[$\onlinecite{ross}$]$ is too large 
to allow one to discriminate between the two behaviors. Further experiments are 
planned to shed light on this issue. 

To conclude, the present study shows that the molecular field induced admixture between CEF levels generates an effective magnetic anisotropy as a plausible mechanism for an energetic selection of the $\psi_2$ state. The proposed model captures a number of key features of the inelastic neutron scattering data, including the opening of a spin gap. Its order of magnitude shows that the proposed mechanism appears as efficient as the ObD scenario, questioning the completeness of the projected pseudospin $1/2$ model \cite{zito,savary,gingras1,gingras2} as a minimal model of \erti. Our study raises the question whether cooperating quantum \cite{zito,savary,gingras1} and thermal \cite{gingras2} order by disorder is the sole or even the principal mechanism for the selection of $\psi_2$ in this material, and whether its advocacy as a rare example of ObD \cite{zito,savary,gingras1,gingras2,zito2} will stand the test of time. On a more positive note, it seems plausible that quantum fluctuations {\it and} anisotropy induced by CEF admixing cooperate to select $\psi_2$ in \erti. Conversely, one might ask whether their {\it competition} might be responsible in part for some of the perplexing properties of \ersn\,\cite{ersn}.

\acknowledgements

We acknowledge E. Lhotel, L. Jaubert and P. McClarty for fruitful discussions, as well as F. Damay for a careful reading of the manuscript. We also thank S. Turc (cryogeny group at ILL) for his technical help while setting up the magnet and the dilution fridge. Research at the Perimeter Institute for Theoretical Physics is supported by the Government of Canada through Industry Canada and by the Province of Ontario through the Ministry of Economic Development \& Innovation.


\section*{Supplemental  Material}

\section{Mean field model}

Our mean field study follows the approach of Ref. \cite{clarty}; it is based on the following Hamiltonian for rare earth (R) moments $\vec{J}_i$ at site $i$ of the pyrochlore lattice:
\begin{equation}
{\cal H} = {\cal H}_{\mbox{CEF}} + \frac{1}{2}\sum_{i,j} \vec{J}_i \cdot {\cal K}_{i,j} \cdot \vec{J}_j 
\label{hh}
\end{equation}
In this expression,  ${\cal H}_{\mbox{CEF}}=\sum_{n,m} B_{nm} O_{nm}$ is written in terms of $O_{nm}$ Stevens operators \cite{wybourne,jensen} with $B_{20} = 616$ K, $B_{40} = 2850$ K, 
$B_{43} = 795$ K, $B_{60} = 858$ K,$B_{43} = -494$ K, $B_{66} = 980$ K, in Weybourne conventions. These parameters lead to CEF excitations at 6.2, 7.5, 13.8, 44.9, 47.7, 52.6 and 69.6 meV which compare well with the first levels determined experimentally through neutron scattering at 6.4, 7.3 and 15.4 meV \cite{champion,shirai,bertin}.

${\cal K} ={\cal J}+{\cal D}$ is the sum of the anisotropic exchange tensor ${\cal J}$ and of the ${\cal D}$ dipolar interaction, limited to the contribution of the nearest-neighbors. Various conventions have been used to define ${\cal K}_{i,j}$ \cite{clarty,savary,zito,thompson,malkin}. Here, ${\cal K}_{i,j}$ is defined in the $(\vec{a},\vec{b},\vec{c})$ frame linked with a R-R bond \cite{malkin}:
\begin{eqnarray*}
\vec{J}_i \cdot {\cal K}_{i,j} \cdot \vec{J}_j  &=& \sum_{\mu,\nu=x,y,z} J_i^{\mu} 
\left( 
{\cal K}_a a_{ij}^{\mu} a_{ij}^{\nu} +  
{\cal K}_b  b_{ij}^{\mu} b_{ij}^{\nu} \right.\\
& & \left.
+ {\cal K}_c  c_{ij}^{\mu} c_{ij}^{\nu} 
\right) J_j^{\nu} +  {\cal K}_4 \sqrt{2}~\vec{b}_{ij}.(\vec{J}_i \times \vec{J}_j)
\end{eqnarray*}
Considering for instance the pair of \er\, ions at $\vec r_1=(1/4,3/4,0)a$ and $\vec r_2=(0,1/2,0)a$, where $a$ is the cubic lattice constant, we define the local bond frame as:  $\vec{a}_{12} = (0,0,-1)$, $\vec{b}_{12} = 1/\sqrt{2} (1,-1,0)$ and $\vec{c}_{12} = 1/\sqrt{2} (-1,-1,0)$. This Hamiltonian, written in terms of bond-exchange constants, has the advantage to provide a direct physical interpretation of the different parameters. Note that ${\cal K}_4$ is an anti-symmetric exchange constant (Dzyaloshinskii-Moriya like), while ${\cal K}_{a,b,c}$ are symmetric terms. Owing to the form of the dipolar interaction, we have :
\begin{equation}
{\cal D}_{i,j} = D_{nn} \left(\vec{a}_{ij}\vec{a}_{ij} + \vec{b}_{ij}\vec{b}_{ij} - 2 \vec{c}_{ij}\vec{c}_{ij} \right)
\end{equation}
with $D_{nn} = \frac{\mu_o}{4\pi}\frac{(g_J \mu_B)^2}{r^3_{nn}}$ and where $r_{nn}$ is the nearest neighbour distance in the pyrochlore lattice. If we combine them with the anisotropic exchange constants $({\cal J}_a,{\cal J}_b,{\cal J}_c)$, we obtain : 
\begin{eqnarray*}
{\cal K}_a  &=& {\cal J}_a + D_{nn} \\
{\cal K}_b  &=& {\cal J}_b + D_{nn} \\ 
{\cal K}_c  &=& {\cal J}_c - 2  D_{nn}
\end{eqnarray*}
As is usual in mean-field approximations, a self-consistent treatment is carried out to solve the problem: starting from a random configuration for the $\langle \vec{J}_j \rangle$, the contribution to $\cal H$ at site $i$ is diagonalized in the Hilbert space of the \er\, magnetic moment defined by the $\left\{ | J_z \rangle \right\}, J_z=-15/2,...,15/2$ basis vectors. This yields the energies $E_{i,\mu}$ and the wave functions $\vert \phi_{i,\mu} \rangle$. The updated magnetic moments, $\langle \vec{J}_i  \rangle'$, at each step of the iteration procedure, are given by: 
\begin{equation}
\langle \vec{J}_i  \rangle' =  \sum_{\mu} \frac{e^{-E_{i,\mu}/k_B T}}{Z} \langle  \phi_{i,\mu}  | \vec{J}_i | \phi_{i,\mu} \rangle
\end{equation}
with
\begin{equation}
Z =\sum_{\mu} \exp{-E_{i,\mu}/k_B T}
\end{equation}
These are then used to proceed to site $j$, and this is repeated until convergence.


\section{Relation with quantum pseudo-spin half models}

This anisotropic exchange Hamiltonian can be rewritten in terms of couplings between the spin components of a pseudospin $1/2$ defined in subspace spanned by the ground CEF doublet states \cite{savary}:
\begin{eqnarray*}
{\cal H} &=& \sum_{i,j} {\sf J}_{zz} {\sf S}^z_i {\sf S}^z_j - {\sf J}_{\pm} \left({\sf S}^+_i {\sf S}^-_j + {\sf S}^-_i {\sf S}^+_j + \right) \\
& &
+ {\sf J}_{\pm\pm} \left(\gamma_{ij} {\sf S}^+_i {\sf S}^+_j + \gamma^*_{ij} {\sf S}^-_i {\sf S}^-_j \right) \\
& &
+ {\sf J}_{z \pm} \left[ {\sf S}_i^z \left( \zeta_{ij} {\sf S}^+_j + \zeta^*_{ij} {\sf S}^-_j\right) + i \leftrightarrow j \right] 
\end{eqnarray*} 
$({\sf J}_{\pm\pm},{\sf J}_{\pm},{\sf J}_{z\pm},{\sf J}_{zz})$ is the set of effective exchange parameters. Note that the ``sanserif" notation refers to local bases. The states of this pseudospin $1/2$ span the ground CEF wavefunctions doublet, using the relation :
\begin{equation}
g_J \vec{{\sf J}} = g \vec{{\sf S}}~~~\mbox{or}~~~~ \vec{{\sf J}} = \lambda \vec{{\sf S}}
\end{equation}
For the pyrochlores, of Fd$\bar{3}$m space group, $\lambda=\frac{g}{g_J}$ matrix is diagonal and takes the form:
\begin{equation}
\lambda = \left(
\begin{array}{ccc}
\lambda_{\perp} & & \\
 & \lambda_{\perp} & \\
& & \lambda_z
\end{array}
\right)
\end{equation}
We finally obtain the following relations:
\begin{eqnarray*}
{\sf J}_{zz}      & = & \lambda_z^2 ~\frac{{\cal K}_a-2{\cal K}_c-4{\cal K}_4}{3} \\
{\sf J}_{\pm}    & = & -\lambda_{\perp}^2 ~\frac{2{\cal K}_a-3{\cal K}_b-{\cal K}_c+4{\cal K}_4}{12} \\
{\sf J}_{z\pm}   & = & \lambda_{\perp}~\lambda_z ~\frac{{\cal K}_a+{\cal K}_c-{\cal K}_4}{3 \sqrt{2}} \\
{\sf J}_{\pm \pm} & = & \lambda_{\perp}^2 ~\frac{2{\cal K}_a+3{\cal K}_b-{\cal K}_c+4{\cal K}_4}{12}
\end{eqnarray*}
and, conversely:
\begin{eqnarray*}
{\cal K}_a & = & \frac{4}{3}~\frac{{\sf J}_{\pm\pm} - {\sf J}_{\pm }}{\lambda_{\perp}^2} + \frac{4 \sqrt{2}}{3} \frac{{\sf J}_{z\pm}}{\lambda_{\perp}\lambda_z} + \frac{1}{3} \frac{{\sf J}_{zz}}{\lambda_z^2} \\
{\cal K}_b & = & 2~\frac{{\sf J}_{\pm\pm} + {\sf J}_{\pm }}{ \lambda_{\perp}^2} \\
{\cal K}_c & = & \frac{2}{3}~\frac{-{\sf J}_{\pm\pm} + {\sf J}_{\pm }}{ \lambda_{\perp}^2} + \frac{4 \sqrt{2}}{3} \frac{{\sf J}_{z\pm}}{\lambda_{\perp}\lambda_z} - \frac{2}{3} \frac{{\sf J}_{zz}}{\lambda_z^2} \\
{\cal K}_4 & = & \frac{2}{3}~\frac{{\sf J}_{\pm\pm} - {\sf J}_{\pm }}{ \lambda_{\perp}^2} - \frac{\sqrt{2}}{3} \frac{{\sf J}_{z\pm}}{\lambda_{\perp}\lambda_z} - \frac{1}{3} \frac{{\sf J}_{zz}}{\lambda_z^2}
\end{eqnarray*}


\section{Details about the magnetic structure}
\label{mag_struc}

The description of the possible magnetic structures in \erti\ is based on the symmetry 
analysis performed in the Fd$\bar{3}$m space group for a ${\bf k}=0$ propagation vector. 
Different structures are possible, belonging to different irreducible representations. 
As explained in the main text, \erti\, undergoes a transition towards an antiferromagnetic
 N\'eel phase below $T_N \approx1.2$~K \cite{Blote69, Harris98, Siddharthan99}. This ordered 
phase corresponds to the $\psi_2$ basis vector of the $\Gamma_5$ irreducible representation. 
The magnetic moment at site i is defined in a local frame $(\vec{a}_i,\vec{b}_i,\vec{e}_i)$ 
given in Table \ref{table1}. Owing to symmetry, the moments in the $\psi_2$ configuration are 
of the form $(x,x,y), (-x,-x,y), (-x,x,-y), (x,-x,-y)$. The experiment favors $y=2x$, hence 
a non-collinear structure, in which the magnetic moments are perpendicular to the local 
$\langle 111 \rangle$ axes. In the general case, the projection of the moment along the 
CEF axis is:
\begin{equation}
m_z = \frac{2x -y}{\sqrt{3}}
\end{equation}
while the angle formed with the XY plane is: 
\begin{equation}
\cos \theta  = \frac{1}{\sqrt{3}}\frac{(2x -y)}{2x^2+y^2}
\end{equation}
Assuming classical magnetic moments of equal amplitude $m=\sqrt{2x^2+y^2}$ on 
the different sites, and calculating the coupling matrices ${\cal K}$, it is possible to calculate 
analytically the classical energy per spin in the $\psi_2$ state: 
\begin{eqnarray}
E_c &= &\frac{1}{2}\left(2 {\cal K}_4 + 4 {\cal K}_a- 3 {\cal K}_b-5{\cal K}_c \right.\\
& & \left. +3 (2 {\cal K}_4+{\cal K}_b-{\cal K}_c) \cos 2\theta \right) m^2
\end{eqnarray}
In the XY case, $y=2x$:
\begin{equation}
E_c= (2 {\cal K}_a- 3 {\cal K}_b- {\cal K}_c) m^2
\end{equation}

\begin{table}[h]
\begin{tabular}{ccccc}
\hline
Site & 1 & 2 & 3 & 4 \\
CEF axis $\vec{e}_i$ & (1,1,-1) & (-1,-1,-1) & (-1,1,1) & (1,-1,1) \\
\hline
Position          & ($\frac{1}{4}$, $\frac{3}{4}$, 0) & (0, $\frac{1}{2}$, 0) & (0, $\frac{3}{4}$,$\frac{1}{4}$) & ($\frac{1}{4}$, $\frac{1}{2}$, $\frac{1}{4}$) \\
$\vec{a}_i$       & (-2,1,-1) & (2,-1,-1) & (2,1,1) & (-2,-1,1) \\
$\vec{b}_i$       & (0,1,1) & (0,-1,1) & (0,1,-1) & (0,-1,-1) \\
\hline
\end{tabular}
\caption{$(\vec{a}_i,\vec{b}_i,\vec{e}_i)$ frame for the different sites of a tetrahedron.}
\label{table1}
\end{table}

\section{Spin dynamics in the RPA}

The spin dynamics is calculated using the random phase approximation (RPA) \cite{jensen}. Spin excitations are constructed on the basis of the transitions between the mean-field states $|\phi_{i,\mu} \rangle$ of energy $E_{i,\mu}$. These transitions occur between the set of CEF modes, perturbed or split by the molecular field. 

Because of the interactions between the magnetic moments, these transitions acquire a dispersion. Two quantities are essential, namely the matrix element of $\vec{J}$ that connects the two states $|\phi_{i,\mu}\rangle$ and $|\phi_{i,\nu}\rangle$:
\begin{equation}
\vec{u}_{i \mu \nu} = \langle \phi_{i \mu} \vert \vec{J}_{i} - \langle \vec{J}_{i} \rangle \vert \phi_{i \nu}\rangle
\end{equation}
and the energy of the transition $(E_{i,\mu}-E_{i,\nu})$.\\

Following the generalized susceptibility approach \cite{jensen,rotter,kao}, it is useful to introduce labels associated with a transition $s=(i,\mu \rightarrow \nu),$ as well as matrices $A_s$ and energies $\Delta_s$ so that: 
\begin{equation}
\Delta_s = E_{i,\nu} - E_{i,\mu}
\end{equation}
and:
\begin{equation}
A_s^{ab} = \left(n_{i,\mu}-n_{i,\nu} \right)~u^{\alpha}_{i \mu \nu}~u^{\beta}_{i \nu \mu}
\end{equation}
with $n_{i,\mu} = \exp{(-E_{i,\mu}/k_{\rm B}T})/Z$. The RPA generalized  spin-spin susceptibility tensor $\tilde \chi_{s,s'}(\vec{Q},\omega)$ is a solution of \cite{jensen,kao}:
\begin{equation}
{\tilde \xi}_{s}~\delta_{s,s'} = 
\sum_{s''}
\left[ \delta_{s,s''}~-{\tilde \xi}_{s}{\cal \tilde K}_{s,s''}(\vec{Q})
\right] {\tilde \chi}_{s'',s'}
\label{rpa}
\end{equation}
Here, $ {\tilde \xi}_{s}(\omega)$ is the 3 $\times$ 3 single-ion spin-spin susceptibility: 
\begin{equation}
{\tilde \xi}_{s}(\omega)^{\alpha \beta} = 
\frac{A_s}
{\hbar\omega+i0^+  - \Delta_s}
\label{eq16}
\end{equation}
The spin-spin correlation function ${\cal S}(\vec{Q},\omega)$ is then given by:
\begin{eqnarray}
{\cal S}(\vec{Q},\omega) = \frac{|F(\vec{Q})|^2}{1-\exp{-(\hbar \omega / k_B T)}} 
\sum_{\alpha, \beta=x,y,z} ~\nonumber \\
\left(\delta_{\alpha \beta}-\frac{Q^\alpha Q^\beta}{Q^2}\right) 
\mbox{Im}~\left[\sum_{s,s'}{\tilde \chi}^{\alpha \beta}_{s,s'}(\vec{Q},\omega)e^{i \vec{Q}(\vec r_i-\vec r_j)}\right] 
\end{eqnarray}
where $\vec{r}_i$ denotes the position of the ion at site $i$ and $F(\vec{Q})$ is the magnetic form factor of the \er magnetic ions. \\

We proceed by rewriting the RPA Eq. (\ref{rpa}) using Eq. (\ref{eq16}):
\begin{equation}
A_s \delta_{s,s'} = \sum_{s''} \left(  (\hbar\omega-\Delta_s) I \delta_{s,s''} -A_s {\cal K}_{s,s''} \right) {\tilde \chi}_{s'',s'}  \\
\label{eq17}
\end{equation}
Owing to the definition of $A_s$, it is possible to show that this matrix has just one non zero eigenvalue $\alpha_s$:
\begin{equation}
\alpha_s =\sum_a A_s^{aa} =  \left(n_{i,\mu}-n_{i,\nu} \right)~\sum_a |u^{a}_{i \mu \nu}|^2
\end{equation}
while the corresponding (normalized) eigenvector is nothing but $\vec{U}_s=\frac{\vec{u}_s}{|u_s|}$. Multiplying Eq. (\ref{eq17}) on the left and on the right by $U^+_s$ and $U_{s'}$ respectively, we get:
\begin{eqnarray}
\alpha_s \delta_{s,s'} &=&\sum_{s''} \left( (\hbar\omega-\Delta_s) \delta_{s,s''} - \alpha_s U_s^+ {\cal K}_{s,s''} U_{s''} \right) \nonumber\\
& & \times \left( U_{s''}^+{\tilde \chi}_{s'',s'} U_{s'} \right). \nonumber
\end{eqnarray}
We are then left with an inversion problem to determine $U_{s''}^+{\tilde \chi}_{s'',s'} U_{s'}$ and then ${\tilde \chi}_{s'',s'}$. From this equation, it is also worth noting that the pole of the susceptibility are given by the eigenvalues of the matrix $R$: 
\begin{equation}
R_{s,s''} = \Delta_s\delta_{s,s''} + \alpha_s U_s^+ {\cal K}_{s,s''} U_{s''} 
\label{matriceR}
\end{equation}

\section{Determination of the coupling constants : fitting procedure}

\begin{figure}[t]
\centerline{
\includegraphics[width=8cm]{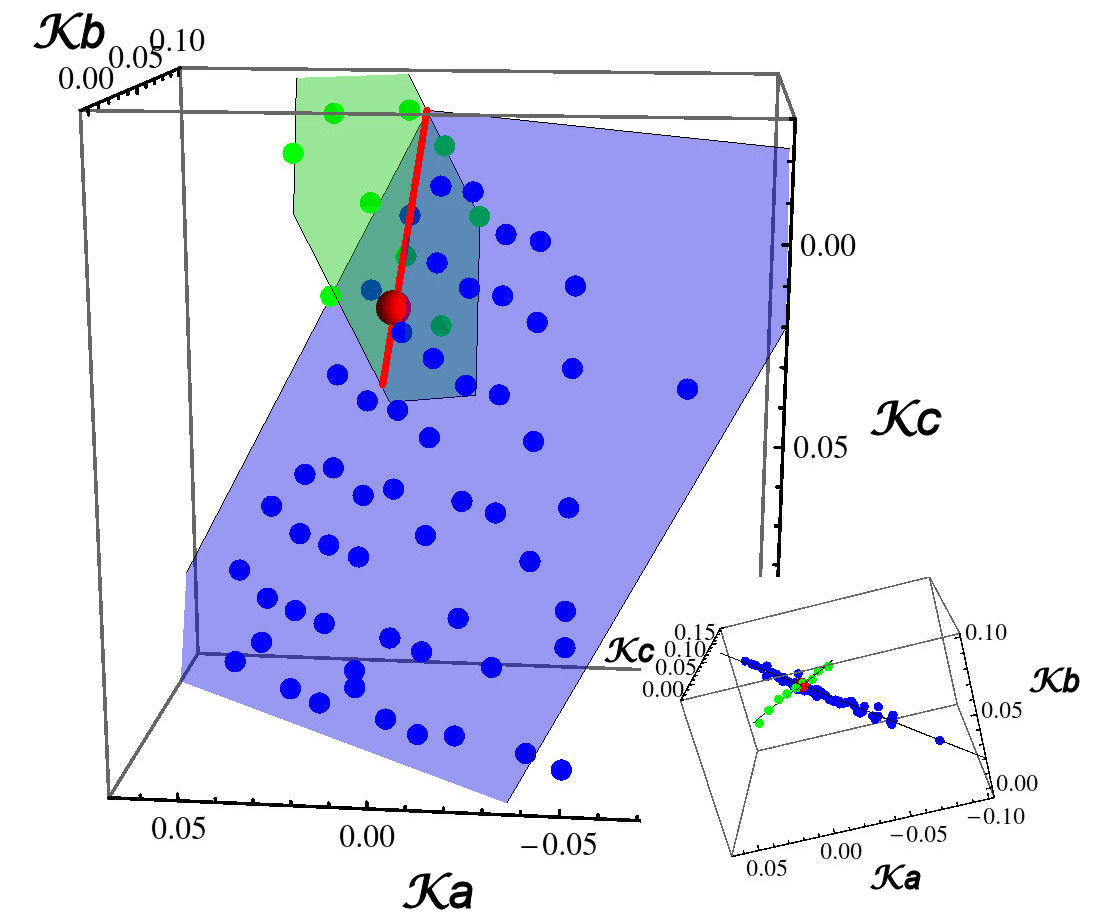}}
\caption{(color online): Best sets of coupling constants $(\mathcal{K}_a,\mathcal{K}_b,\mathcal{K}_c)$. The blue (resp. green) points, obtained from the $H=0$~T (resp 2.5~T) inelastic neutron scattering data, lie around the blue (resp. green) plane, which is better evidenced in the inset. The red line is the intersection of the two planes, and the red point the best fit (see text).}
\label{fig-fit}
\end{figure}

\begin{figure}[t]
\centerline{
\includegraphics[width=6cm]{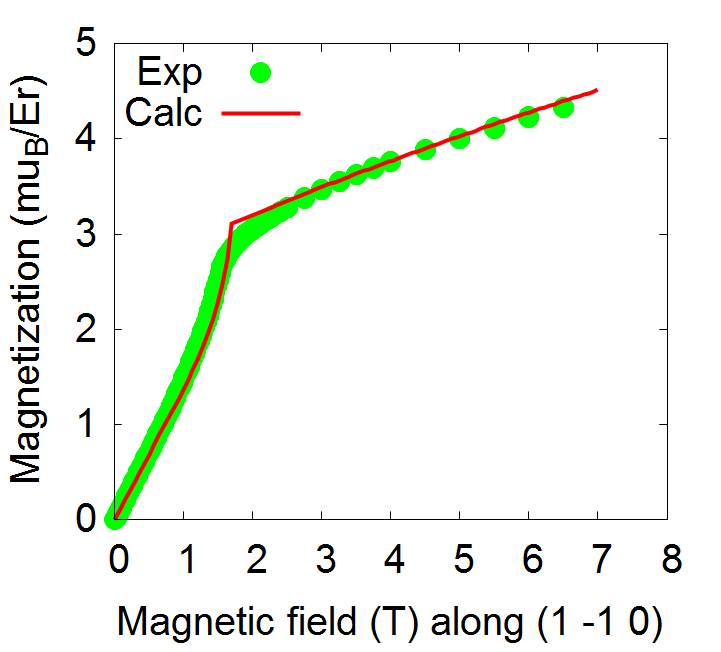}}
\caption{(color online): Magnetization curve calculated with the RPA and using the INS-fitted parameters. The $M(H)$ data is taken from Refs. $[$\onlinecite{bonville,elsa2}$]$. 
}
\label{fig-mh}
\end{figure}

A fitting procedure in a four parameters space may be hazardous. Reducing the parameter space often requires a combination of different experimental techniques giving complementary information as well as a detailed characterization of the system in function of external parameters, such as the temperature or the magnetic field. Ref. \cite{savary} claims that Er$_2$Ti$_2$O$_7$ has a small Dzyaloshinskii-Moriya interaction $\mathcal{K}_4\ll \mathcal{K}_{a,b,c}$. Although it may slightly improve the agreement between simulations and experimental data, we checked that such a contribution has a weak effect on the excitation spectrum. In what follows, we therefore consider $\mathcal{K}_4=0$. Exhaustive calculations of the scattering function $S(\mathbf{Q},\omega)$ have been performed as a function of the remaining parameters $\mathcal{K}_{a,b,c}$, assuming equipopulated $\psi_2$ magnetic domains at zero magnetic field and only one domain at $H=2.5$~T. Experimental data were fitted by matching the location of the maximum neutron intensity in several directions. The spectral weight repartition has also been taken into account, checking that numerical intensities roughly matche the experimental ones. All the considered parameters giving a good agreement are reported in blue (resp. green) for $H=0$~T (resp. 2.5~T) in Figure \ref{fig-fit}. Interestingly, while many sets of parameters capture the zero field data, the ones obtained at $H=2.5$~T appear to be much restrictive: only a very small portion of parameter space accounts for the corresponding experimental dispersions. 
This also points out very simple linear correlations between parameters (which is better evidenced in the inset): the relevant coupling constants are contained into a plane, thus reducing the parameter space to a surface. Different relations between the coupling constants (or surfaces) are obtained considering the zero ($\mathcal{K}_c = 1.90(2)\mathcal{K}_a - 2.53(5)\mathcal{K}_b + 0.218(2)$) or the finite magnetic field data ($\mathcal{K}_c = -2.0(3)\mathcal{K}_a - 4.4(6)\mathcal{K}_b + 0.37(5)$). 
Combining these two latter relations reduces the relevant parameter space to a line represented in red in the Fig. \ref{fig-fit}, whose parametric equations is:
\begin{eqnarray}
\mathcal{K}_a & = & 0.138 \mathcal{K}_c\\
\mathcal{K}_b & = & 0.085 - 0.291 \mathcal{K}_c\\
\mathcal{K}_c & = & \mathcal{K}_c.
\end{eqnarray}
Along this line, the best set of parameters, fitting all the experimental data, is finally obtained for $(\mathcal{K}_a,\mathcal{K}_b,\mathcal{K}_c,\mathcal{K}_4) = (0.003(5),0.075(5),0.034(5),0.000(5))$~K (red point in Fig. \ref{fig-fit}). We finally checked that this set of parameters allows one to capture the magnetization curve measured at $T=110$~mK (see Fig. \ref{fig-mh}) \cite{petrenko,elsa2,bonville}, except near the quantum critical field driven transition at 1.6 T \cite{ruff,cao}, where the mean-field theory is not expected to be accurate. 

\section{Time Of Flight Experiments}

\begin{figure}[t]
\centerline{
\includegraphics[width=8cm]{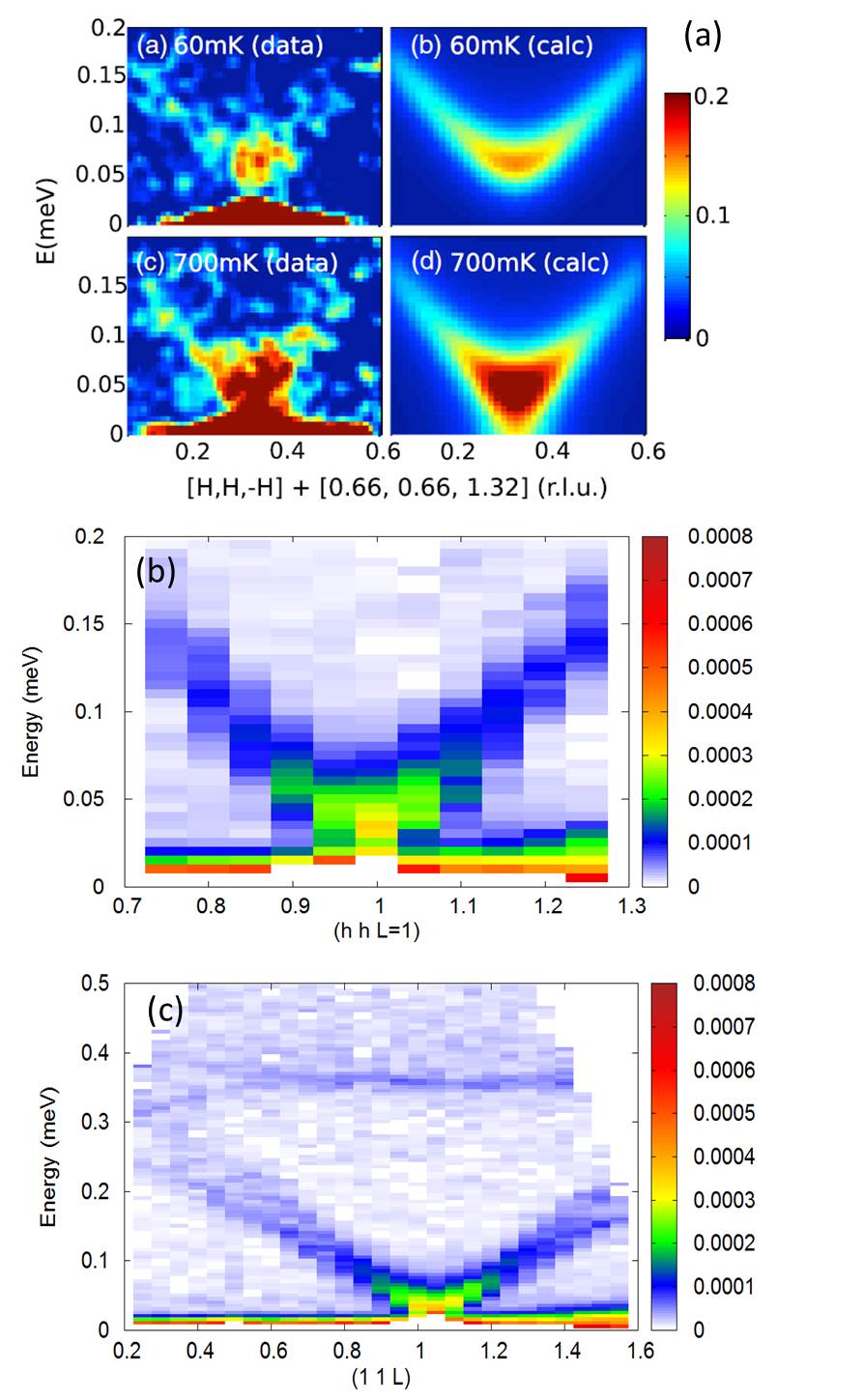}}
\caption{(color online): (a) corresponds to TOF data by Ross {\it et al.} showing the filling of the spin gap with temperature at 50 an 700 mK, together with reconstructed fits \cite{ross}. (b) and (c) shows the data obtained at 50 mK in the present work along $(hh1)$ and $(11L)$ respectively with an incident wavelength of 8.5 \AA\, providing an energy resolution of 20 $\mu$eV.
}
\label{fig-gap}
\end{figure}

Data were collected on the IN5 time-of-flight instrument (ILL) in its single-crystal set up. Measurements were carried out with an incident neutron wavelength of 6 and 8.5 \AA, in zero field and under an applied magnetic field of 1.5 and 2.5 T along $[1,-1,0]$ at 50 mK. Note that Ref.~[\onlinecite{ruff}] shows data along $[0,0,L]$ while the main body of Ref.~[\onlinecite{savary}] shows in field data only. Figure \ref{fig-gap} shows selected maps from the 8.5 \AA\, data, allowing one to observe the spin gap at the zone center $(1,1,1)$. Those data were obtained by integrating the full 4D data-set in narrow Q-range by steps of 0.05 ``rlu", and then reassembled to produce the maps. The data by Ross {\it et al.} is shown for comparison. Note that the latter was obtained with a resolution of 12 $\mu$eV \cite{ross}, while it was 20 $\mu$eV in the present experiment. The two acoustic-like modes coming from different magnetic domains are clearly visible along $(11L)$, along with the depletion of the intensity below the spin gap.

\begin{thebibliography}{99}
\bibitem{bnm} The crystal field is modeled by the following coefficients: $B_{20} = 616$ K, $B_{40} = 2850$ K, 
$B_{43} = 795$ K, $B_{60} = 858$ K,$B_{43} = -494$ K, $B_{66} = 980$ K, in Weybourne conventions.
\end{thebibliography}
\end{document}